\documentclass[12pt]{article}
\usepackage{amsfonts}
\usepackage{amsmath}
\usepackage{amssymb}
\usepackage{setspace}
\usepackage{graphicx}%
\begin{document}

\title{ \begin{flushright}
{\small CECS-PHY-06/06 }
\end{flushright}
\vskip 1.0cm Simple compactifications and Black p-branes in Gauss-Bonnet and Lovelock Theories}
\author{Gaston Giribet$^{1,2}$, Julio Oliva$^{3,4}$, Ricardo
Troncoso$^{4}$\footnote{e-mails: gaston-at-df.uba.ar,
juliooliva-at-cecs.cl, ratron-at-cecs.cl}  \and $^{1}${\small
Departamento de Fisica, Universidad de Buenos Aires} \and {\small
Ciudad Universitaria, Pabellon I (1428), Buenos Aires, Argentina.}
\and $^{2}${\small Instituto de Fisica de La Plata, Universidad
Nacional de La Plata,} \and {\small C.C. 67, 1900, La Plata,
Argentina.} \and $^{3}${\small Departamento de Fisica, Universidad
de Concepcion, Casilla 160-C, Concepcion, Chile.} \and
$^{4}${\small Centro de Estudios Cientificos (CECS), Casilla 1469,
Valdivia, Chile.}} \maketitle

\begin{abstract}
We look for the existence of asymptotically flat simple
compactifications of the form $\mathcal{M}_{D-p}\times T^{p}$ in
$D$-dimensional gravity theories with higher powers of the
curvature. Assuming the manifold $\mathcal{M}_{D-p}$ to be
spherically symmetric, it is shown that the Einstein-Gauss-Bonnet
theory admits this class of solutions only for the pure
Einstein-Hilbert or Gauss-Bonnet Lagrangians, but not for an
arbitrary linear combination of them. Once these special cases
have been selected, the requirement of spherical symmetry is no
longer relevant since actually any solution of the pure Einstein
or pure Gauss-Bonnet theories can then be toroidally extended to
higher dimensions. Depending on $p$ and the spacetime dimension,
the metric on $\mathcal{M}_{D-p}$ may describe a black hole or a
spacetime with a conical singularity, so that the whole spacetime
describes a black or a cosmic $p$-brane, respectively. For the
purely Gauss-Bonnet theory it is shown that, if
$\mathcal{M}_{D-p}$ is four-dimensional, a new exotic class of
black hole solutions exists, for which spherical symmetry can be
relaxed.
Under the same assumptions, it is also shown that simple
compactifications acquire a similar structure for a wide class of
theories among the Lovelock family which accepts this toroidal
extension.
The thermodynamics of black $p$-branes is also
discussed, and it is shown that a thermodynamical analogue of the
Gregory-Laflamme transition always occurs regardless the spacetime
dimension or the theory considered, hence not only for General
Relativity.

Relaxing the asymptotically flat behavior, it is also shown that exact black
brane solutions exist within a very special class of Lovelock theories.

\end{abstract}

\vskip 1.0cm \tableofcontents
\newpage

\setcounter{equation}{0}
\section{Introduction}

The metric theory of gravity consisting in second order equations
of motion for the Riemann tensor and leading, besides, to a
conserved stress-tensor for the matter fields is unique and is the
quoted Lovelock theory of gravity \cite{Lovelock:1971yv}. As it
was early pointed out by Lanczos \cite{Lanczos-1938}, this theory
does not lead to classical modifications to the four-dimensional
Einstein's theory of general relativity, though actually differs
from that in higher dimensions. In the generic case, such
differences correspond to short-distance modifications to the
Einstein theory and, even though these become negligible at large
scales, actually lead to important corrections to the short scale
physics. Perhaps, the black hole physics is the most celebrated
example of this; for which both the thermodynamical and
geometrical features turn out to be substantially modified by the
inclusion of additional terms in the action of Lovelock theory.

The Lovelock Lagrangian density in $D$ dimensions is
\begin{equation}
\mathcal{L}=\sum_{n=0}^{N}a_{n}\mathcal{L}_{n}\ , \label{lov1}%
\end{equation}
where $2N=D-2$ (for even dimensions $D$) while $2N=D-1$ (for odd dimensions
$D$). In (\ref{lov1}), $a_{n}$ are arbitrary constants which represent the
coupling of the terms in the Lagrangian, and $\mathcal{L}_{n}$ is given by%
\begin{equation}
\mathcal{L}_{k}=\frac{1}{2^{k}}\sqrt{-g}\delta_{j_{1}...j_{2k}}^{i_{1}%
...i_{2k}}R_{\qquad i_{1}i_{2}}^{j_{1}j_{2}}...R_{\qquad i_{2k-1}i_{2k}%
}^{j_{2k-1}j_{2k}}\ . \label{lov2}%
\end{equation}
Here ${R^{\mu}{}_{\nu\rho\gamma}}$ is the Riemann tensor, $g$ is the
determinant of the metric $g_{\mu\nu}$ and $\delta_{j_{1}...j_{2k}}%
^{i_{1}...i_{2k}}$ is the generalized Kronecker delta of order $2k$. Then, the
action reads%
\begin{equation}
I=\int d^{D}x\mathcal{L}\ .
\end{equation}

With this notation, the Lagrangian up to second order is given by the sum of
three terms; namely
\begin{align*}
\mathcal{L}_{0}  &  =\sqrt{-g}\ ,\\
\mathcal{L}_{1}  &  =\frac{1}{2}\sqrt{-g}\delta_{j_{1}j_{2}}^{i_{1}i_{2}%
}R_{\qquad i_{1}i_{2}}^{j_{1}j_{2}}=\sqrt{-g}R\ ,\\
\mathcal{L}_{2}  &  =\frac{1}{4}\sqrt{-g}\delta_{j_{1}j_{2}j_{3}j_{4}}%
^{i_{1}i_{2}i_{3}i_{4}}R_{\qquad i_{1}i_{2}}^{j_{1}j_{2}}R_{\qquad i_{3}i_{4}%
}^{j_{3}j_{4}}=\sqrt{-g}(R_{\mu\nu\rho\sigma}R^{\mu\nu\rho\sigma}-4R_{\mu\nu
}R^{\mu\nu}+R^{2})\ ,
\end{align*}
For dimensions $D=5$ and $D=6$ the Lovelock Lagrangian is a linear
combination of the Einstein-Hilbert term $\mathcal{L}_{1}$ and the
often called Gauss-Bonnet term $\mathcal{L}_{2}$, which receives
such a name because it corresponds to the Euler density in four
dimensions. The theory in dimensions higher than five could also
include the cubic Lagrangian $\mathcal{L}_{3}$, whose physical
implications were early studied by M\"{u}ller-Hoissen
\cite{Mueller-Hoissen:1985mm} and have been revisted recently in
\cite{Dehghani:2005vh} and \cite{Dehghani:2006dh}. Lovelock terms
can also been understood in the context of BRST cohomology
\cite{Cnockaert:2005jw}.

The complexity of the Lovelock theory, basically due to higher
order terms in the curvature tensor as well as the plethora of
coupling constants, makes that the task of finding analytical
exact solutions for this turns out to be a highly non-trivial
problem (see, for instance, Ref. \cite{Radu:2006mb}). However, as
we will see below, for certain class of solutions to exist the
coefficients $a_{n}$ have to be fine-tuned in a precise form. This
choice corresponds to be the same as requiring the theory to have
a unique maximally symmetric vacuum with a fixed cosmological
constant as in Ref. \cite{BH-Scan}. It turns out that for such
cases the solutions can be found in a closed form. Let us discuss
an example of this: Consider the case of the Einstein-Gauss-Bonnet
theory in five dimensions, whose equations
of motion take the form%
\[
\Lambda g_{\mu\nu}+\mathcal{\beta}\left(  R_{\mu\nu}-\frac{1}{2}Rg_{\mu\nu
}\right)  +\alpha\left(  2RR_{\mu\nu}-4R_{\mu\rho}R_{\quad\nu}^{\rho}%
-4R_{\rho\delta}R_{\quad\mu\nu}^{\rho\delta}\right.
\]%
\[
+2R_{\mu\rho\delta\gamma}R_{\nu }^{\quad\rho\delta\gamma}-\left.
\frac{1}{2}g_{\mu\nu}(R_{\rho\delta\gamma\lambda}R^{\rho\delta
\gamma\lambda}-4R_{\rho\delta}R^{\rho\delta}+R^{2})\right)  =0\ .
\]
Then, if a simple compactification\footnote{By simple
compactification we mean a space which is a solution of the vacuum
field equations in the absence of
Kaluza-Klein gauge fields and with a constant dilaton.} of the form $\mathcal{M}%
_{4}\times S^{1}$ is considered, with $\mathcal{M}_{4}$
representing a four-dimensional asymptotically flat solution with
spherical symmetry; then, one is unavoidably led to the conclusion
that the only way for obtaining a non-trivial solution is that of
setting one of the coefficients $\alpha$ or $\beta$ to zero.
Namely, the only possibilities turn out to be $\Lambda =\alpha=0$
and $\Lambda=\beta=0$, i. e., only for the pure Einstein-Hilbert
or Gauss-Bonnet Lagrangians, but not for an arbitrary linear
combination of them. In Section II, this is discussed in detail
and, furthermore, it is also shown that the same effect occurs for
solutions of the form $\mathcal{M}_{D-p}\times T^{p}$, for
arbitrary $D$ and $p$. Depending on $p$ and the spacetime
dimension, the metric on $\mathcal{M}_{D-p}$ may describe a black
hole or a spacetime with a conical singularity, so that the whole
spacetime describes a black or a cosmic $p$-brane, respectively.
For the purely Gauss-Bonnet theory it is shown that, if
$\mathcal{M}_{D-p}$ is four-dimensional, a new exotic class of
black hole solutions exists, for which spherical symmetry can be
relaxed. It is also shown that simple compactifications of the
form $\mathcal{M}_{d}\times T^{p}$ in $D=d+p$ acquire a similar
structure for the case of a theory described by a Lagrangian given
by an arbitrary single term $\mathcal{L}_{k}$ in Eq. (\ref{lov1}).
The seven-dimensional case is instructive because it captures the
whole structure; and it is discussed in Section III. Section IV is
devoted to the discussion of the general case, i.e., for a
Lagrangian given by $\mathcal{L}_{k}$ in$\ D$ dimensions and with
arbitrary $p$. The thermodynamics of black $p$-branes is discussed
in Section V, where it is shown that a thermodynamical analogue of
the Gregory-Laflamme transition always occurs regardless the
spacetime dimension or the theory considered. Section VI is
devoted to the summary and to the discussion of the exotic case as
well as to show that the \ asymptotically flat behavior can be
relaxed, and exact black brane solutions with a warp factor exist
within the class of theories discussed in Ref. \cite{BH-Scan}.

\emph{Note added:} The same day this paper was sent to
arXiv:hep-th, the paper \cite{Kastor:2006vw} appeared in the same
database, which contains some overlap with our results.

\section{Einstein-Gauss-Bonnet Lagrangian: selecting the theories}

In this section, we look for asymptotically flat simple compactifications of
the form $\mathcal{M}_{D-p}\times T^{p}$ in $D$-dimensional
Einstein-Gauss-Bonnet theory. By demanding the manifold $\mathcal{M}_{D-p}$ to
be spherically symmetric, it is shown that the Einstein-Gauss-Bonnet theory
admits this class of solutions only for the pure Einstein-Hilbert or
Gauss-Bonnet Lagrangians, but not for an arbitrary linear combination of them.
Once these special cases have been selected, the requirement of spherical
symmetry is no longer relevant since actually any solution of the pure
Einstein or pure Gauss-Bonnet theories can then be cylindrically extended to
higher dimensions. Depending on $p$ and the spacetime dimension $D$, the
metric on $\mathcal{M}_{D-p}$ may describe a black hole solution or a
spacetime with a conical singularity, so that the whole spacetime describes a
black or a cosmic $p$-brane, respectively. For the purely Gauss-Bonnet theory,
it is shown that, if $\mathcal{M}_{D-p}$ is four-dimensional ($D-p=4$), a new
exotic class of black hole solutions does exist, for which spherical symmetry
can be relaxed.

Then, we are interested in solutions of the form%
\begin{equation}
ds^{2}=d\tilde{s}_{D-p}^{2}+\sum_{n=1}^{p}R_{0}^{(n)}d\phi_{n}^{2}\ ,
\label{entire}%
\end{equation}
where $\mathcal{M}_{D-p}$ with metric $d\tilde{s}_{D-p}^{2}$ is assumed to
have spherical symmetry and the element $\sum_{n=1}^{p}R_{0}^{(n)}d\phi
_{n}^{2}\ $ denotes the flat metric on the $T^{p}$ and $R_{0}^{\left(
n\right)  }$ is the radius of the $n$-th $S^{1}$ factor. The spacetime
indices of the manifold $\mathcal{M}_{D-p}\times T^{p}$ will be split
according to $\alpha_{1},...,\alpha_{D-p}$ for $\mathcal{M}_{D-p}$ and
$\phi_{1},...,\phi_{p}$ for $T^{p}$. Analogously, the tangent space indices of
the manifold $\mathcal{M}_{D-p}\times T^{p}$ will be denoted by capital
letters in the begin of the alphabet i.e. $A,B,...$ and will be split as
$\mu_{1},...,\mu_{D-p}$ for $\mathcal{M}_{D-p}$ and $i_{n}=i_{1},...,i_{p}$
for $T^{p}$.

In the next subsection, in order to present the idea and sketch the procedure,
we first discuss the particular case $D=7$ and $p=1$. In the following
subsection, we analyze the case for any dimension $D$ and arbitrary $p$.
Besides, notice that for $p=D$, the entire manifold is $T^{D}$, while for
$p=D-1$, the entire manifold is $\mathbb{R\times}T^{D-1}$. Thus in what
follows, we consider $p<D-1$.

The details of our results can be enormously simplified, and thus explicitly
done, by the use of differential forms. The Lovelock action then reads%
\begin{equation}
I=\int\sum_{n=0}^{[\frac{D-1}{2}]}\alpha_{n}\mathcal{L}^{n}~,
\end{equation}
with%
\begin{equation}
\mathcal{L}^{n}=\epsilon_{a_{1}...a_{2n}a_{2n+1}...a_{D}}R^{a_{1}a_{2}%
}...R^{a_{2n-1}a_{2n}}e^{a_{2n+1}}...e^{a_{D}}~,
\end{equation}
and where $e^{a}$ and $R^{ab}$ stand for the vielbein and the
curvature two-form. Here, $[x]$
 stands for the integer part of $x$. The field equations obtained through the variation with respect to
the vielbein and the spin connection then read (see e.g. \cite{HDG})%
\begin{align}
\sum_{n=0}^{[\frac{D-1}{2}]}\left(  D-2n\right)  \alpha_{n}\mathcal{E}%
_{a}^{n}  &  =0\ ,\label{Ea}\\
\sum_{n=1}^{[\frac{D-1}{2}]}n\left(  D-2n\right)  \alpha_{n}\mathcal{E}%
_{ab}^{n}  &  =0\ , \label{Eab}%
\end{align}
respectively, and where%
\begin{align}
\mathcal{E}_{a}^{n}  &  =\epsilon_{ab_{1}...b_{D-1}}R^{b_{1}b_{2}%
}...R^{b_{2n-1}b_{2n}}e^{b_{2n+1}}...e^{b_{D-1}}~,\\
\mathcal{E}_{ab}^{n}  &  =\epsilon_{abc_{1}...c_{D-2}}R^{c_{1}c_{2}%
}...R^{c_{2n-1}c_{2n}}T^{c_{2n+1}}e^{c_{2n+2}}...e^{c_{D-1}}~.
\end{align}
In what follows, we assume that the torsion two-form $T^{a}$ vanishes, so that
Eq. (\ref{Eab}) is trivially satisfied.

The vielbeins and the curvature corresponding to (\ref{entire}) are given by%
\begin{equation}
e^{A}=\left\{
\begin{array}
[c]{c}%
e^{\mu_{n}}=\tilde{e}^{\mu_{n}}\\
e^{j_{n}}=R_{0}^{\left(  j_{n}\right)  }d\phi^{j_{n}}%
\end{array}
\right.  ~, \label{VielbeinMxT}%
\end{equation}
and%
\begin{equation}
R^{AB}=\left(
\begin{array}
[c]{cc}%
R^{\mu_{m}\nu_{n}} & R^{\mu_{m}j_{n}}\\
R^{j_{m}\mu_{n}} & R^{\phi_{m}\phi_{n}}%
\end{array}
\right)  =\left(
\begin{array}
[c]{cc}%
\tilde{R}^{\mu\nu} & 0\\
0 & 0
\end{array}
\right)  ~. \label{CurvatureMxT}%
\end{equation}

Now, let us move to the seven dimensional case.

\subsection{A working example: Einstein-Gauss-Bonnet theory in $D=7$ and
$p=1$}

Let us consider the Einstein-Gauss-Bonnet theory in $D=7$ dimensions, whose
field equations read%
\begin{equation}
\epsilon_{ABCDEFG}\left(  7\alpha_{0}e^{B}e^{C}e^{D}e^{E}e^{F}e^{G}%
+5\alpha_{1}R^{BC}e^{D}e^{E}e^{F}e^{G}+3\alpha_{2}R^{BC}R^{DE}e^{F}%
e^{G}\right)  =0\text{ }.\label{EGB7}%
\end{equation}
In order to have an asymptotically flat solution, it is necessary to require
that the coefficient $\alpha_{0}$ that multiplies volume term vanishes. This
can be easily seen as follows. Since Eq. (\ref{EGB7}) can be factorized as%
\begin{equation}
\gamma_{2}\ \epsilon_{ABCDEFG}\left(  R^{BC}+\gamma_{1}e^{B}e^{C}\right)
\left(  R^{DE}+\gamma_{0}e^{D}e^{E}\right)  e^{F}e^{G}=0~,\label{ecfactorized}%
\end{equation}
with $7\alpha_{0}=\gamma_{0}\gamma_{1}\gamma_{2}$, it is apparent then that
asymptotically flat solutions can only be found either if $\gamma_{0}=0$ or
$\gamma_{1}=0$, which in turn means that $\alpha_{0}$ must vanish. Thus, the
suitable theory possessing asymptotically flat solutions has the following
field equations%
\begin{equation}
\mathcal{E}_{A}=\epsilon_{ABCDEFG}\left(  3\alpha_{2}R^{BC}R^{DE}e^{F}%
e^{G}+5\alpha_{1}R^{BC}e^{D}e^{E}e^{F}e^{G}\right)  =0\ .\label{ecf2}%
\end{equation}
Considering a spacetime with geometry given by $\mathcal{M}_{6}\times S^{1}$,
and splitting the indices as $A=\left\{  \mu,1\right\}  $, the field
equations, $\mathcal{E}_{A}=0$ in (\ref{ecf2}) become%

\begin{align}
\mathcal{E}_{\mu} &  =\epsilon_{\mu\nu\lambda\rho\sigma\tau1}\left(
3\alpha_{2}\tilde{R}^{\nu\lambda}\tilde{R}^{\rho\sigma}\tilde{e}^{\tau
}+10\alpha_{1}\tilde{R}^{\nu\lambda}\tilde{e}^{\rho}\tilde{e}^{\sigma}%
\tilde{e}^{\tau}\right)  e^{1}=0\ ,\label{epsmucero}\\
\mathcal{E}_{1} &  =\epsilon_{1\nu\lambda\rho\sigma\tau\mu}\left(  3\alpha
_{2}\tilde{R}^{\nu\lambda}\tilde{R}^{\rho\sigma}\tilde{e}^{\tau}\tilde{e}%
^{\mu}+5\alpha_{1}\tilde{R}^{\nu\lambda}\tilde{e}^{\rho}\tilde{e}^{\sigma
}\tilde{e}^{\tau}\tilde{e}^{\mu}\right)  =0\ .\label{epsicero}%
\end{align}
Requiring spherical symmetry on $\mathcal{M}_{6}$, by virtue of
the generalization of Birkhoff's theorem \cite{Zegers:2005vx} and
\cite{DFr}, Eq. (\ref{epsmucero}) implies that the metric
$d\tilde{s}_{6}$ on $\mathcal{M}_{6}$ is the one found by Boulware
and Deser \cite{Boulware:1985wk}. On the other hand, considering
the combinations
$e^{\mu}\mathcal{E}_{\mu}-e^{1}\mathcal{E}_{1}=0$, and $e^{\mu
}\mathcal{E}_{\mu}-2e^{1}\mathcal{E}_{1}=0$ one obtains additional
constraints
on the geometry of $\mathcal{M}_{6}$%
\begin{align}
\alpha_{1}\ \left(  \epsilon_{1\nu\lambda\rho\sigma\tau\mu}\tilde{R}%
^{\nu\lambda}\tilde{e}^{\rho}\tilde{e}^{\sigma}\tilde{e}^{\tau}\tilde{e}^{\mu
}\right)  \ e^{1} &  =0~,\label{Coalpha1}\\
\alpha_{2}\ \left(  \epsilon_{1\nu\lambda\rho\sigma\tau\mu}\tilde{R}%
^{\nu\lambda}\tilde{R}^{\rho\sigma}\tilde{e}^{\tau}\tilde{e}^{\mu}\right)
\ e^{1} &  =0~,\label{Calpha2}%
\end{align}
which is equivalent to say that each term in Eq. (\ref{epsicero}) must vanish separately.

For the generic case where $\alpha_{1}$ and $\alpha_{2}$ are different from
zero, the constraints (\ref{Coalpha1}) and (\ref{Calpha2}) turn out to be too
strong, since they imply that spacetime must be flat. In other words, in this
case the constraints are satisfied by the Boulware-Deser solution only if the
mass vanishes. Therefore, in order to circumvent this obstruction for the
existence of nontrivial solutions, one has to require that either
$\alpha_{1}$ or $\alpha_{2}$ vanish.

In the case $\alpha_{2}=0$, one recovers Einstein's theory and the remaining
equation (\ref{Coalpha1}) (which just means the vanishing of the Ricci scalar)
is no longer a constraint and generates no incompatibility since it just
corresponds to the trace of the field equation (\ref{epsmucero}).

The remaining possibility is to consider $\alpha_{1}=0$, i.e., the gravity
theory described by the purely Gauss-Bonnet term. Analogously, in this case
the remaining equation (\ref{Calpha2}) is not a constraint since it is just
the trace of the field equation (\ref{epsmucero}) and the incompatibility is
then removed.

In sum, we have shown that requiring the existence of an asymptotically flat
solution of the form $\mathcal{M}_{6}\times S^{1}$ we obtain that the volume
term must be absent ($\alpha_{0}=0$), and assuming the manifold $\mathcal{M}%
_{6}$ to be spherically symmetric, it is shown that the Einstein-Gauss-Bonnet
theory admits this class of solutions only for the pure Einstein-Hilbert or
Gauss-Bonnet Lagrangians, but not for an arbitrary linear combination of them.
Furthermore, once these special cases have been selected, one may notice that
the requirement of spherical symmetry is no longer relevant since actually any
solution of the six-dimensional pure Einstein or pure Gauss-Bonnet theories
can then be cylindrically extended to seven dimensions.

\subsection{Einstein-Gauss-Bonnet theory for arbitrary $D$ and $p$}

The case of arbitrary $p$ in $D$ dimensions is performed by the
straightforward generalization of the results found in the previous subsection.

The field equations for Einstein-Gauss-Bonnet theory now read%
\[
\epsilon_{ab_{1}...b_{D-1}}\left(
D\alpha_{0}e^{b_{1}}...e^{b_{D-1}}+\left( D-2\right)
\alpha_{1}R^{b_{1}b_{2}}e^{b_{3}}...e^{B_{D-1}}\right.
\]
\[\left.
+\left( D-4\right)
\alpha_{2}R^{b_{1}b_{2}}e^{b_{3}}...e^{b_{D-1}}\right)  =0\ ,
\]
and again, requiring the existence of asymptotically flat solutions forces one
to choose $\alpha_{0}=0$. Then, the suitable theory possessing asymptotically
flat solutions reads%
\begin{equation*}
(D-4)\alpha_{2}\epsilon_{AB_{1}...B_{D-1}}R^{B_{1}B_{2}}R^{B_{3}B_{4}}%
e^{B_{5}}...e^{B_{D-1}}
\end{equation*}
\begin{equation}
+(D-2)\alpha_{1}\epsilon_{AB_{1}...B_{D-1}}%
R^{B_{1}B_{2}}e^{B_{3}}e^{B_{4}}e^{B_{5}}...e^{B_{D-1}}=0\ .
\end{equation}
We consider spacetimes of the form $\mathcal{M}_{D-p}\times T^{p}$, where
$T^{p}$ stands for the $p$-dimensional flat torus. Indices now split so
that greek ones $\mu,\nu,\lambda$ run along $\mathcal{M}_{D-p}$, and latin indices
$i,j,k$ along $T^{p}$. Then, the field equations now split as
\begin{equation}%
\begin{array}
[c]{c}%
\mathcal{E}_{\mu_{1}}=\epsilon_{\mu_{1}\mu_{2}...\mu_{D-p}i_{1}...i_{p}%
}\left[  \binom{D-5}{p}(D-4)\alpha_{2}\tilde{R}^{\mu_{2}\mu_{3}}\tilde{R}%
^{\mu_{4}\mu_{5}}\tilde{e}^{\mu_{6}}...\tilde{e}^{\mu_{D-p}}e^{i_{1}%
}...e^{i_{p}}\right. \\
\left.  +\binom{D-3}{p}(D-2)\alpha_{1}\tilde{R}^{\mu_{2}\mu_{3}}\tilde{e}%
^{\mu_{4}}...\tilde{e}^{\mu_{D-p}}e^{i_{1}}...e^{i_{p}}\right]
\end{array}
=0~, \label{epsmuzrodarb}%
\end{equation}%
\begin{equation}%
\begin{array}
[c]{c}%
\mathcal{E}_{i_{1}}=\epsilon_{i_{1}\mu_{1}...\mu_{D-p}i_{2}...i_{p}}\left[
\binom{D-5}{p-1}(D-4)\alpha_{2}\tilde{R}^{\mu_{1}\mu_{2}}\tilde{R}^{\mu_{3}%
\mu_{4}}\tilde{e}^{\mu_{5}}...\tilde{e}^{\mu_{D-p}}e^{i_{2}}...e^{i_{p}%
}\right. \\
\left.  +\binom{D-3}{p-1}(D-2)\alpha_{1}\tilde{R}^{\mu_{1}\mu_{2}}\tilde
{e}^{\mu_{3}}...\tilde{e}^{\mu_{D-p}}e^{i_{2}}...e^{i_{p}}\right]
\end{array}
=0~, \label{epsizerodarb}%
\end{equation}
where the decomposition of the vielbein and the curvature as in Eqs.
(\ref{VielbeinMxT}) and (\ref{CurvatureMxT}) has been used.

Requiring spherical symmetry on $\mathcal{M}_{D-p}$, and using the
generalization of Birkhoff's theorem \cite{Zegers:2005vx} and
\cite{DFr}, the field equation along $\mathcal{M}_{D-p}$,
(\ref{epsmuzrodarb}) implies that the metric of
$\mathcal{M}_{D-p}$ corresponds to the Boulware-Deser solution
\cite{Boulware:1985wk}. Then, suitable linear combinations of the
trace field
equations allow give rise to the following constraints%
\begin{equation}
\alpha_{1}\epsilon_{i_{1}\mu_{1}...\mu_{D-p}i_{2}...i_{p}}\tilde{R}^{\mu
_{1}\mu_{2}}\tilde{e}^{\mu_{3}}...\tilde{e}^{\mu_{D-p}}e^{i_{2}}...e^{i_{p}%
}=0\ ,\label{ca1d}%
\end{equation}%
\begin{equation}
\alpha_{2}\epsilon_{i_{1}\mu_{1}...\mu_{D-p}i_{2}...i_{p}}\tilde{R}^{\mu
_{1}\mu_{2}}\tilde{R}^{\mu_{3}\mu_{4}}\tilde{e}^{\mu_{5}}...\tilde{e}%
^{\mu_{D-p}}e^{i_{2}}...e^{i_{p}}=0\ ,\label{ca2d}%
\end{equation}
implying that each term in Eq. (\ref{epsizerodarb}) must vanish separately.

Again, for the generic case where $\alpha_{1}$ and $\alpha_{2}$ are different
from zero, the constraints (\ref{ca1d}) and (\ref{ca2d}) annihilate the mass
of the Boulware-Deser solution implying that spacetime must be flat. Hence, in
order to obtain nontrivial solutions it is necessary to require that either
$\alpha_{1}$ or $\alpha_{2}$ vanish. In the case $\alpha_{2}=0$, the remaining
equation (\ref{ca1d}) is no longer a constraint since it corresponds to the
trace of the field equation (\ref{epsmuzrodarb}). For the remaining
possibility, $\alpha_{1}=0$, Eq. (\ref{ca2d}) is not a constraint since it
becomes the trace of the field equation (\ref{epsmuzrodarb}) and the generic
incompatibility is thus removed.

Therefore, it has been shown that requiring the existence of an
asymptotically flat solution of the form $\mathcal{M}_{D-p}\times
T^{p}$, the volume term must be absent ($\alpha_{0}=0$), and
assuming the manifold $\mathcal{M}_{D-p}$ to be spherically
symmetric, the Einstein-Gauss-Bonnet theory was shown to admit
this class of solutions only for the pure Einstein-Hilbert or pure
Gauss-Bonnet cases, but not for an arbitrary linear combination of
them. Having selected these special cases, the requirement of
spherical symmetry can be dropped out since actually any solution
of the $(D-p)$-dimensional pure Einstein or pure Gauss-Bonnet
theories can then be toroidally extended to $D$
dimensions\footnote{Let us mention here that cylindrical
extensions of  four-dimensional solutions in Einstein-Gauss-Bonnet
theory were also discussed in references \cite{otras},
\cite{otras3}, \cite{otras2}.}.

\subsection{Summary and extension to theories with higher powers in the
curvature}

We have shown that the existence of non trivial toroidal extensions of
asymptotically flat and spherically symmetric solutions for the
Einstein-Gauss-Bonnet theory is only achieved for the cases where the
Lagrangian is selected in the form
\[
\mathcal{L}=\mathcal{L}^{(k)}\ \ ,\text{\ for }k\geq1\ ,
\]
which corresponds to the pure Gauss-Bonnet Lagrangian
($\mathcal{L}^{(2)}=\epsilon RRe...e$), or for the pure
Einstein-Hilbert case ($\mathcal{L}^{(1)}=\epsilon Re...e$), and
that in these cases the requirement of spherical symmetry can be
dropped out because for this special class of theories any
solution of the $(D-p)$-dimensional pure Einstein or pure
Gauss-Bonnet theories can then be toroidally extended to $D$
dimensions.

One may then naturally wonder whether this results extend to theories with
higher powers in the curvature.

Based on the previous results, we consider the theory described by a
Lagrangian given by the dimensional continuation of the Euler density of
dimension $2k$, i.e.,
\begin{equation}
I^{(k)}=\frac{\kappa_{D,k}}{\left(  D-2k\right)  }\int\epsilon_{B_{1}%
...B_{2k}B_{2k+1}...B_{D}}R^{B_{1}B_{2}}...R^{B_{2k-1}B_{2k}}e^{B_{2k+1}%
}...e^{B_{D}}\ . \label{actionk}%
\end{equation}

This action possesses well behaved spherically symmetric black hole solutions
\cite{BH-Scan} for $k<\left[  \frac{D}{2}\right]  $ with metric%
\begin{equation}
ds_{d}^{2}=-\left(  1-\left(  \frac{2G_{k}m}{r^{D-2k-1}}\right)  ^{\frac{1}%
{k}}\right)  dt^{2}+\frac{dr^{2}}{\left(  1-\left(  \frac{2G_{k}m}{r^{D-2k-1}%
}\right)  ^{\frac{1}{k}}\right)  }+r^{2}d\Omega_{D-2}^{2}~, \label{BHdk}%
\end{equation}
where $m$ is the mass and the gravitational constant\footnote{The usual Newton
constant is related with this one through $8\pi G_{Newton}=(d-2)\Omega
_{d}G_{k=1}$, where $\Omega_{d}$ is the volume of unit sphere in $d$
dimensions, being $\Omega_{d}=2\pi^{\frac{d+1}{2}}/\Gamma(\frac{d+1}{2})$.}
$G_{k}$ is related to the coupling constant in the action $\kappa_{D,k}$ by%
\begin{equation}
\kappa_{D,k}=\frac{1}{2\left(  D-2\right)  !\Omega_{D-2}G_{k}}\text{~}.
\label{kGrelation}%
\end{equation}
For $D=2k+1$, (\ref{BHdk}) describes a conical singularity.

One can then see that these theories are special in the same sense described
above, since any of their solutions for a given dimension can be toroidally
extended to higher dimensions within the same theory. Therefore the class of
black hole solutions (\ref{BHdk}) can be extended to black $p$-branes. It is
then natural to see whether there exists a Gregory-Laflamme-like transition
for these kind of objects within this class of theories which are quite
different from General Relativity.

It is also shown here that for some particular cases, the
toroidally extended solutions may describe cosmic strings as well
as black strings with exotic topology in the transverse section.
This last kind of exotic black $p$-branes objects belong to a
completely different class of solutions, since they do not
correspond to the toroidal extension of the black holes found in
\cite{BH-Scan}.

In the next section we discuss the seven-dimensional case since it is a simple
and good representative that captures the whole structure present in the
general case.

\section{Scanning the seven-dimensional case with extended objects}

The field equations for the seven-dimensional class of theories described by
the action $I^{(k)}$ in Eq. (\ref{actionk}) are given by%
\begin{equation}
\epsilon_{AB_{1}...B_{2k}B_{2k+1}...B_{6}}R^{B_{1}B_{2}}...R^{B_{2k-1}B_{2k}%
}\overset{6-2k}{\overbrace{e^{B_{2k+1}}...e^{B_{6}}}}=0\ .
\end{equation}
Let us consider a spacetime of the form $\mathcal{M}_{7-p}\times T^{p}$, whose metric is
of the form (\ref{entire}). Before entering into the details we summarize how
the structure depends on the values of $k$ and $p$:

\bigskip

\begin{itemize}
\item $p<6-2k$ : Black $p$-brane, where $\mathcal{M}_{7-p}$ is a black hole of the form
(\ref{BHdk}).

\item $p=6-2k$ : Cosmic $p$-brane, since in this case $\mathcal{M}_{7-p}$ is given by
(\ref{BHdk}) which describes a conical singularity.

\item $p=7-2k$ : Exotic black $p$-brane, where $\mathcal{M}_{7-p}$ is given by a new
kind of black hole geometry for which the Euler density vanishes.

\item $6>p\geq8-2k$ : The manifold $\mathcal{M}_{7-p}$ is
arbitrary.
\end{itemize}

\bigskip

In the following subsections we describe in detail these results.

\subsection{Case $k=1$: The Einstein theory}

The toroidal extensions of the Einstein-Hilbert theory are
well-known and we discuss them here just for completeness. For
$p\geq6$ the whole spacetime is locally flat, while for $p=5$ the
manifold $\mathcal{M}_{2}$ must be locally flat, which in two
dimensions is trivially equivalent to say that its Euler density
vanishes, i.e., $\varepsilon_{2}\left(  \mathcal{M}\right)  =0$.
For $p=4$, one recovers the cosmic brane, since in this case the
manifold $\mathcal{M}_{3}$ solves the three-dimensional Einstein
equations without cosmological constant whose spherically
symmetric solutions have a conical singularity. The remaining
cases correspond to $p<4$, where the solutions are Black
$p$-branes where $\mathcal{M}_{7-p}$ is endowed with the
Schwarzschild metric.

\subsection{Case $k=2$: The pure Gauss-Bonnet theory}

For $k=2$, the equations of motion are given by%
\begin{equation}
\epsilon_{ABCDEFG}R^{BC}R^{DE}e^{F}e^{G}=0\ . \label{d7k2}%
\end{equation}

\subsubsection{$p<2$: Black string and black hole}

The black hole (\ref{BHdk}) is obviously recovered for $p=0$.

For $p=1$ the geometry describes a Black string. In this case, the field
equations split as%
\begin{align}
\epsilon_{\mu\nu\lambda\rho\sigma\tau}\tilde{R}^{\nu\lambda}\tilde{R}%
^{\rho\sigma}\tilde{e}^{\tau}  &  =0\ ,\label{d7k2p1mu0}\\
\epsilon_{\mu\nu\lambda\rho\sigma\tau}\tilde{R}^{\nu\lambda}\tilde{R}%
^{\rho\sigma}\tilde{e}^{\tau}\tilde{e}^{\mu}  &  =0\ ,
\end{align}
where the second Eq. gives no extra restrictions since it is just the trace of
(\ref{d7k2p1mu0}). The problem has then been reduced to solve the same
equations for $\mathcal{M}_{6}$. Therefore, as any solution of the theory with $k=2$ can
be cylindrically extended, a black string is obtained from to the cylindrical
extension of the six-dimensional black hole (\ref{BHdk}) with $k=2$.

\subsubsection{$p=2$: Cosmic membrane}

For $p=2$ the field equations split according to%
\begin{align}
\epsilon_{\mu\nu\lambda\rho\tau}\tilde{R}^{\nu\lambda}\tilde{R}^{\rho\tau}  &
=0\ ,\label{d7k2p2mu0}\\
\epsilon_{\mu\nu\lambda\rho\tau}\tilde{R}^{\nu\lambda}\tilde{R}^{\rho\tau
}\tilde{e}^{\mu}  &  =0\ ,
\end{align}
and again the second equation is the trace of the equation for $\mathcal{M}_{5}$. The
problem has been reduced to the one of finding a solution for the purely
Gauss-Bonnet theory in five dimensions. Since the spherically symmetric
solution is a spacetime with a conical singularity, its toroidal extension
corresponds to a Cosmic membrane.

\subsubsection{$p=3$: The exotic black three-brane}

In the case of $p=3$ a very interesting phenomenon occurs, since in this case
the projection of the field equations along $\mathcal{M}_{4}$ is trivially satisfied
($\mathcal{E}_{\mu}\equiv0$) because at least one of the indices along $T^{3}$
lies in a curvature. The remaining field equation then reads%

\begin{equation}
\varepsilon_{4}\left(  \mathcal{M}_{4}\right)  :=\epsilon_{\mu\nu\lambda\rho
}\tilde{R}^{\mu\nu}\tilde{R}^{\lambda\rho}=0\ .\label{Euleren4}%
\end{equation}
This means that we have a single scalar equation for the four-manifold $\mathcal{M}_{4}$
which states that its Euler density vanishes. Note that this is a very weak
condition on the geometry of $\mathcal{M}_{4}$, as compared with the equations
for a standard gravity theory on a four-dimensional manifold. An explicit
solution of Eq. (\ref{Euleren4}) that includes a black hole with exotic
topology is presented in Section VI. Therefore its toroidal extension
originates the exotic black $3$-brane.

\subsubsection{$6>p\geq4$: The manifold $\mathcal{M}_{7-p}$ is arbitrary}

In this case the field equations are trivially satisfied $(\mathcal{E}_{\mu
}\equiv0$, $\mathcal{E}_{i}\equiv0)$ since the torus is ``big enough'' to ensure
that at least one of the indices along $T^{p}$ lies always in the curvatures.
Therefore, as the equations of motion are identically solved, one obtains no
restriction on the geometry of $\mathcal{M}_{7-p}$.

\subsection{Case $k=3$: Beyond the Einstein and Gauss-Bonnet theories}

The field equations for $k=3$ read%
\begin{equation}
\epsilon_{ABCDEFG}R^{BC}R^{DE}R^{FG}=0\ .
\end{equation}

In the case of $p=0$, the spherically symmetric solution with
conical singularity is recovered from Eq. (\ref{BHdk}), and no
seven-dimensional toroidal extension of the black holes found in
\cite{BH-Scan} exist for any value of $p$.

\subsubsection{$p=1$: The exotic string}

In the case of $p=1$ the projection of the field equations along $\mathcal{M}_{6}$ is
again trivially satisfied ($\mathcal{E}_{\mu}\equiv0$). The remaining field
equation now reads%

\begin{equation}
\varepsilon_{6}\left(  \mathcal{M}_{6}\right)  :=\epsilon_{\mu\nu
\lambda\rho\sigma\tau}\tilde{R}^{\mu\nu}\tilde{R}^{\lambda\rho}\tilde
{R}^{\sigma\tau}=0~,\label{euleren4mas2}%
\end{equation}
which means that the Euler density of $\mathcal{M}_{6}$ vanishes. A solution for Eq.
(\ref{euleren4mas2}) that includes a black hole with exotic topology is
presented in Section VI, so that its cylindrical extension is the exotic black string.

\subsubsection{$6>p\geq2$: The manifold $\mathcal{M}_{7-p}$ is arbitrary}

In this case the field equations are again trivially satisfied since the
dimension of the torus is large enough. This means that there is no
restriction on the geometry of $\mathcal{M}_{7-p}$.

In the next section the results in seven dimensions are generalized for
arbitrary $D$, $k$ and $p$.

\section{Arbitrary $D,k$ and $p$}

The field equations read%
\begin{equation}
\epsilon_{AB_{1}...B_{2k}B_{2k+1}...B_{D-1}}R^{B_{1}B_{2}}...R^{B_{2k-1}%
B_{2k}}\overset{D-2k-1}{\overbrace{e^{B_{2k+1}}...e^{B_{D-1}}}}=0~.
\label{eqdkparb}%
\end{equation}
Consider a spacetime of the form $\mathcal{M}_{D-p}\times T^{p}$,
with a metric given by (\ref{entire}). The summary of how the
structure depends on $p$ for the $D$-dimensional theory with $k$
curvatures given by (\ref{actionk}) is:

\bigskip

\begin{itemize}
\item $p<D-2k-1$: Black $p$-brane, where $\mathcal{M}_{D-p}$ has a
black hole metric of the form (\ref{BHdk}).

\item $p=D-2k-1$: Cosmic $p$-brane, where $\mathcal{M}_{D-p}$ is
given by (\ref{BHdk}) and describes a spacetime with a conical
singularity.

\item $p=D-2k$: Exotic black $p$-brane, where $\mathcal{M}_{D-p}$
has a new kind of black hole metrics for which the Euler density
vanishes.

\item $D-1>p\geq D-2k+1$: The manifold $\mathcal{M}_{D-p}$ is
arbitrary.
\end{itemize}

\bigskip

The splitting of the fields equations generically is%

\begin{align}
\mathcal{E}_{\mu}  & =\epsilon_{\mu\nu_{1}...\nu_{2k}...\nu_{D-p-1}%
j_{1}...j_{p}}\tilde{R}^{\nu_{1}\nu_{2}}...\tilde{R}^{\nu_{2k-1}\nu_{2k}%
}\overset{D-2k-1}{\overbrace{\underset{D-2k-1-p}{\underbrace{\tilde{e}%
^{\nu_{2k+1}}...\tilde{e}^{\nu_{D-p-1}}}}e^{j_{1}}...e^{j_{p}}}}=0~,\\
\mathcal{E}_{i}  & =\epsilon_{i\nu_{1}...\nu_{2k}\nu_{2k+1}...\nu_{D-p}%
j_{1}...j_{p-1}}\tilde{R}^{\nu_{1}\nu_{2}}...\tilde{R}^{\nu_{2k-1}\nu_{2k}%
}\overset{D-2k-1}{\overbrace{\underset{D-2k-p}{\underbrace{\tilde{e}%
^{\nu_{2k+1}}...\tilde{e}^{\nu_{D-p}}}}e^{j_{1}}...e^{j_{p-1}}}}=0~,
\end{align}

and the analysis goes as follows:

\subsection{$p<D-2k-1$: The black $p$-branes (The problem reduces to find a
black hole for the same theory in $(D-p)$-dimensions)}

For $p=0$ the black hole (\ref{BHdk}) is directly recovered.

For the rest of the allowed range of $p$ the geometry correspond to a black
$p$-brane. In this case, the field equations now split in the following way%
\begin{align}
\epsilon_{\mu\nu_{1}...\nu_{2k}...\nu_{D-p-1}}\tilde{R}^{\nu_{1}\nu_{2}%
}...\tilde{R}^{\nu_{2k-1}\nu_{2k}}\tilde{e}^{\nu_{2k+1}}...\tilde{e}%
^{\nu_{D-p-1}}  &  =0~,\\
\epsilon_{\nu_{1}...\nu_{2k}\nu_{2k+1}...\nu_{D-p-1}\nu_{D-p}}\tilde
{R}^{\nu_{1}\nu_{2}}...\tilde{R}^{\nu_{2k-1}\nu_{2k}}\tilde{e}^{\nu_{2k+1}%
}...\tilde{e}^{\nu_{D-p-1}}\tilde{e}^{\nu_{D-p}}  &  =0~.
\end{align}
The field equations along $T^{p}$ give no extra conditions since they reduce
to the trace of the field equation along $\mathcal{M}_{D-p}$. Therefore the problem
reduces to solve the field equations for the same theory in $\left(
D-p\right)  $-dimensions. Consequently the black $p$-branes are described by
the toroidal extension of the black holes discussed in \cite{BH-Scan}.

\subsection{$p=D-2k-1$: The cosmic $p$-branes}

In this case the field equations split as follows%
\begin{align*}
\epsilon_{\mu\nu_{1}...\nu_{D-p-1}}\tilde{R}^{\nu_{1}\nu_{2}}...\tilde
{R}^{\nu_{D-p-2}\nu_{D-p-1}}  &  =0~,\\
\epsilon_{\nu_{1}...\nu_{D-p-1}\nu_{D-p}}\tilde{R}^{\nu_{1}\nu_{2}%
}...\tilde{R}^{\nu_{D-2p-2}\nu_{D-2p-1}}\tilde{e}^{\nu_{D-p}}  &  =0~,
\end{align*}
so that the equations along $T^{p}$ give again the trace of the equation along
$\mathcal{M}_{2k+1}$. As the problem is reduced the one of finding a solution for the
same theory in $\left(  2k+1\right)  $-dimensions, one can use the spherically
symmetric conical solutions for $\mathcal{M}_{2k+1}$ to obtain the cosmic $p$-branes
from their toroidal extensions.

\subsection{$p=D-2k$: The exotic black $p$-branes with vanishing Euler
density on $\mathcal{M}_{2k}$}

For $p=D-2k$, the projection of the field equations along $\mathcal{M}_{2k}$ is
trivially satisfied ($\mathcal{E}_{\mu}\equiv0$) because the torus $T^{p}$ is
large enough so as at least one of associated indices always lies in a
curvature. A very interesting phenomenon then occurs, since the remaining
field equation implies the vanishing of the Euler density of $\mathcal{M}%
_{2k}$, i.e. :
\begin{equation}
\varepsilon_{2k}\left(  \mathcal{M}_{2k}\right)  :=\epsilon_{\nu_{1}%
...\nu_{D-p}}\tilde{R}^{\nu_{1}\nu_{2}}...\tilde{R}^{\nu_{D-p-1}\nu_{D-p}}=0~.
\label{Euleren2k}
\end{equation}
This is a very weak condition on the geometry of
$\mathcal{M}_{2k}$, since it is just a scalar equation. An
explicit solution for Eq. (\ref{Euleren2k}) including a black hole
with exotic topology is discussed in Section VI. The toroidal
extensions of these black objects generate the exotic black
$p$-branes.

\subsection{$D-1>p\geq D-2k+1$: The manifold $\mathcal{M}_{D-p}$ is arbitrary}

In this case the dimension of the torus is large enough so as to ensure that
the field equations are trivially satisfied, which means that the geometry of
$\mathcal{M}_{D-p}$ is completely unrestricted.

\section{Thermodynamics}

In this section, we study the thermodynamics of the black
$p$-brane solutions that can be obtained from the toroidal
extension of the black holes in Eq.(\ref{BHdk}). The computation
of the mass and the entropy is explicitly performed, and their
thermodynamical stability is briefly discussed for the
microcanonical ensemble, which suggest the extension of the
Gregory-Laflamme instability \cite{Gregory:1993vy} for the class
of theories considered here. We first discuss the particular case
$p=1$ and then describe the thermodynamics for $p<D-2k-1$.

We have shown that for the cases under consideration, the equations of motion
for the whole space $\mathcal{M}_{D-p}\times T^{p}$ in the theory
(\ref{actionk}) with $k$ powers in the curvature reduce to the field equations
of the same theory for the $\left(  D-p\right)  $-dimensional manifold
$\mathcal{M}_{D-p}$. Analogously, it is simple to show that the Euclidean
action evaluated on the black $p$-brane in $D$ dimensions is proportional to
the Euclidean action of the black hole in $\mathcal{M}_{D-p}$. This means that
there is a mapping between \ the thermodynamics of the black $p$-branes and
the thermodynamics of the black holes in $\mathcal{M}_{D-p}$. The
thermodynamics of these black holes was studied in \cite{BH-Scan}, and the
temperature, the mass and the entropy for the black holes in Eq.(\ref{BHdk})
where shown to be given by
\begin{align}
T  &  =\frac{1}{4\pi k_{B}}\frac{\left(  D-2k-1\right)  }{k}\frac{1}{r_{+}%
}~,\label{termobhT}\\
m  &  =\frac{1}{2G_{k}}r_{+}^{D-2k-1}~,\label{termobhM}\\
S_{k}  &  =\frac{2\pi k_{B}}{G_{k}}\frac{k}{\left(  D-2k\right)  }r_{+}%
^{D-2k}~, \label{termobhS}%
\end{align}
respectively. Here $k_{B}$ stands for the Boltzmann constant.

\subsection{The action for the black $p$-brane}

Let us begin with the simplest case $p=1$, though for arbitrary dimension $D$
and with arbitrary $k$. Evaluating the Euclidean action (\ref{actionk}) for
any simple compactification of the form $\mathcal{M}_{D-1}\times S^{1}$ gives%

\begin{align}
I_{D,k}  &  =2\pi
R_{0}\kappa_{D}\int\epsilon_{\mu_{1}...\mu_{2k}\mu
_{2k+1}...\mu_{D-1}\phi}\tilde{R}^{\mu_{1}\mu_{2}}...\tilde{R}^{\mu_{2k-1}%
\mu_{2k}}\tilde{e}^{\mu_{2k+1}}...\tilde{e}^{\mu_{D-1}}\\
&  =\frac{\kappa_{D-1}^{\prime}}{\left(  D-1-2k\right)  }\int\epsilon
_{\mu_{1}...\mu_{2k}\mu_{2k+1}...\mu_{D-1}\phi}\tilde{R}^{\mu_{1}\mu_{2}%
}...\tilde{R}^{\mu_{2k-1}\mu_{2k}}\tilde{e}^{\mu_{2k+1}}...\tilde{e}%
^{\mu_{D-1}}\\
&  =I_{D-1,k}^{\prime}~,
\end{align}
where we have defined
\begin{equation}
\kappa_{D-1}^{\prime}=2\pi R_{0}\kappa_{D}\left(  D-1-2k\right)  ~.
\end{equation}

This means that the Euclidean action for the simple compactification
$\mathcal{M}_{D-1}\times S^{1}$ reduces to the Euclidean action for the same
theory on $\mathcal{M}_{D-1}$ with a modified gravitational constant, which by means of
Eq. (\ref{kGrelation}), is given by\footnote{The subindex $d$ in $G_{d,k}$ stands to explicitly refer to the dimension, which turns out to be very important  in the discussion here; cf. Eq. (\ref{BHdk}).}
\begin{equation}
G_{d,k}^{\prime}=\frac{\left(  d-1\right)  }{\left(  d-2k\right)  }%
\frac{\Omega_{d-1}}{\Omega_{d-2}}\frac{1}{2\pi R_{0}}G_{d+1,k}~.
\end{equation}
Therefore we have shown that the thermodynamics for the black
$p$-branes can be reproduced following the lines of
\cite{BH-Scan}, considering the metric of $\mathcal{M}_{D-1}$ to
be given by (\ref{BHdk}) with the modified gravitational
constant.\footnote{It would also be interesting to analyze the
thermodynamics following alternative approaches, see e.g.
\cite{Abdalla:2001as}, \cite{Cvetic:2001bk}, \cite{Clunan:2004tb}
and references therein, as well as to see the mass loss rates and
lifetimes as in \cite{Rizzo:2006uz}}

The entropy for the black string $\mathcal{M}_{D-1}\times S^{1}$ is then given
by%
\begin{align}
S_{k}^{Bs} &  =2\pi
k_{B}\frac{1}{G_{d,k}^{\prime}}\frac{kr_{+}^{\left(
d-2k\right)  }}{\left(  d-2k\right)  }\\
&  =\frac{4\pi^{2}k_{B}k}{\left(  d-1\right)  }\frac{R_{0}\Omega_{d-2}}%
{\Omega_{d-1}}\frac{r_{+}^{\left(  d-2k\right)  }}{G_{d+1,k}},
\end{align}
where the horizon radius $r_{+}$ can be expressed in term of the
black hole mass $m$ in $\mathcal{M}_{D-1}$, namely
\begin{equation}
r_{+}=\left(  2G_{k,d}m\right)  ^{\frac{1}{d-2k-1}}~.
\end{equation}
Hence, the horizon radius can be written also in terms of the modified
gravitational constant and the string mass since%
\begin{equation}
r_{+}=\left(  2G_{k,d}^{\prime}m_{string}\right)
^{\frac{1}{d-2k-1}}~.
\end{equation}

This implies that the entropy of the black string expressed in terms of the
string mass reads%
\begin{equation}
S_{k}^{Bs}=A_{k}\text{ }(m_{string})^{\frac{D-2k-1}{D-2k-2}},
\end{equation}
where the coefficient $A_{k}$ is given by%
\begin{equation}
A_{k}=k\frac{4\pi k_{B}}{(D-2k-1)}\left(  \frac{2\left(  D-2\right)  }{\left(
D-2k-1\right)  }\frac{\Omega_{D-2}}{\Omega_{D-3}}\frac{G_{D,k}}{2\pi R_{0}%
}\right)  ^{\frac{1}{D-2k-2}}.
\end{equation}

Proceeding in the same way, in the generic case for arbitrary $p$ and $k$, the
entropy of the black $p$-brane of the form $\mathcal{M}_{d}\times T^{p}$ in
$D=d+p$ dimensions for the theory (\ref{actionk}) with $k$ powers in the
curvature is found to be given by%
\begin{equation}
S_{k}^{Bp\text{-}b}=4\pi\kappa_{B}\frac{k}{(D-p-2k)}\text{ }(m_{p\text{-}b}%
)^{\frac{D-2k-p}{D-2k-p-1}}(2A_{p,k}G_{D,k})^{\frac{D-2k-p}{D-2k-p-1}-1}~,
\label{fideos}%
\end{equation}
where $m_{p\text{-}b}$ is the mass of the black $p$-brane, where
the factor $A_{p,k}$ is
\[
A_{p,k}=\frac{(D-2k-p-1)!(D-2)!}{(D-2k-1)!(D-2-p)!}\times\frac{\Omega_{D-2}%
}{\Omega_{D-p-2}Vol(T^{p})}~.
\]
Here $Vol(T^{p})$ corresponds to the volume of the $p$-torus.

It is easy to check that the expression for the entropy above reproduces the
right factors as well as the area law for the Einstein theory which is
recovered in the case $k=1$.

\subsection{Thermodynamical instability: Black holes v/s Black $p$-branes}

Now, we have the suitable tools to study the thermodynamical stability of the
black $p$-branes solutions. A naive, but interesting result can be obtained
comparing the entropy of a $D$-dimensional black hole with the one for a black
$p$-brane in the microcanonical ensemble, i.e., for a fixed mass $m$. In this
case the quotient between both entropies is%
\begin{equation}
\frac{S^{Bp\text{-}b}_{k}}{S^{Bh}_{k}}=\frac{D-2k}{D-2k-p}(2G_{k,D})^a
\text{ }(A_{p,k})^{\frac
{1}{D-2k-p-1}}\text{ }m^b~.\label{kuotien}%
\end{equation}
with $a$ and $b$ given by
\[ a :=
\frac{1}{D-2k-p-1}-\frac{1}{D-2k-1},
\]
\[
b:=  \frac{D-2k-p}{D-2k-p-1}-\frac{D-2k}%
{D-2k-1} \ .
\]

This equation implies the existence of a critical mass $m_{c}$ for
which both entropies agree, $S^{B\text{-}pb}_{k}=S^{Bh}_{k}$, that
explicitly depends on $D$, $k$ and $p$. The critical mass defined
the point where the thermodynamic transition occurs, since for
$m>m_{c}$ the black $p$-brane has an entropy greater than the one
of the black hole, meaning that in this case the black $p$-brane
is thermodynamically favoured, and in this sense stable unlike the
black hole. The converse is obtained for $m<m_{c}$. This
transition suggests that a thermodynamic analogue of the
Gregory-Laflamme instability should also exist beyond General
Relativity, at least for the class of theories considered here
\footnote{It is worth pointing out that, even for black holes in
Gauss-Bonnet theories the stability possesses a fairly different
behavior as compared with the Schwarzschild solution
\cite{Dotti:2004sh}, \cite{Neupane:2003vz}, \cite{Gleiser:2005ra},
\cite{Dotti:2005sq}.}. It is worth pointing out that when the
ratio $S^{Bp\text{-}b}_{k}/S^{Bh}_{k}$ is expressed in terms of
the horizon radius, the scaling of such quantity agrees with the
one for the Einstein theory, even though the entropies do not
follow the area law.

\section{Discussion and summary}

Asymptotically flat simple compactifications of the form
$\mathcal{M}_{D-p}\times T^{p}$ were shown to exist for the class
of theories described by the action (\ref{actionk}). If the
manifold $\mathcal{M}_{D-p}$ is assumed to be spherically
symmetric, it was shown that the Einstein-Gauss-Bonnet theory
admits this class of solutions only for the pure Einstein-Hilbert
or Gauss-Bonnet Lagrangians, but not for an arbitrary linear
combination of them. Once these special cases have been selected,
the requirement of spherical symmetry is no longer relevant since
actually any solution of the pure Einstein or pure Gauss-Bonnet
theories can then be toroidally extended to higher dimensions.
Depending on $p$ and the spacetime dimension, the metric on
$\mathcal{M}_{D-p}$ may describe a black hole or a spacetime with
a conical singularity, so that the whole spacetime describes a
black or a cosmic $p$-brane, respectively. Under the same
assumptions, it was also shown that simple compactifications
acquire a similar structure for the whole class of theories
defined by the action (\ref{actionk}).

The thermodynamics of black $p$-branes was also discussed, and it was shown
that a thermodynamical analogue of the Gregory-Laflamme transition should be
expected to occur regardless the value of $k$ and the spacetime dimension, and
hence not only for General Relativity.

A new class of exotic black $p$-branes exist for $p=D-2k$, for which the
manifold $\mathcal{M}_{2k}$ possesses a metric that makes the Euler density to vanish.
As we have seen, the case when $p=D-2k$ tell us that the Euler density of the
$2k$-dimensional $M$ manifold must vanish. A black hole metric of this kind
for $\mathcal{M}_{2k}$ can be found as follows:

Consider an ansatz of the form%
\begin{equation}
d\tilde{s}^{2}=-f^{2}\left(  r\right)  dt^{2}+\frac{dr^{2}}{f^{2}\left(
r\right)  }+r^{2}d\Sigma_{\gamma}^{2}~,
\end{equation}
where $d\Sigma_{\gamma}^{2}$ stands for the line element of a manifold with
constant curvature given by $\gamma$. In the four dimensional case the
equation
\[
\varepsilon_{4}\left(  \mathcal{M}_{4}\right)  :=\epsilon_{\mu\nu\lambda\rho
}\tilde{R}^{\mu\nu}\tilde{R}^{\lambda\rho}=0~,
\]
has the following solution:$\ $%

\begin{equation}
ds^{2}=-\left(  \pm\sqrt{2Cr+B}+\gamma\right)  dt^{2}+\frac{dr^{2}}{\left(
\pm\sqrt{2Cr+B}+\gamma\right)  }+r^{2}d\Sigma_{\gamma}^{2}~,
\end{equation}
for certain integration constants $B$ and $C$, with $\Sigma$ a
2-dimensional manifold of constant curvature $\gamma$. This
solution describes a black hole in the case for negative $\gamma$
and for the branch with the ``plus'' sign.

In the even dimensional case the equation
\begin{equation}
\varepsilon_{2k}\left(  \mathcal{M}_{2k}\right)  :=\epsilon_{\nu_{1}%
...\nu_{D-p}}\tilde{R}^{\nu_{1}\nu_{2}}...\tilde{R}^{\nu_{D-p-1}\nu_{D-p}}=0~,
\end{equation}
admits the following solution%
\begin{equation}
d\tilde{s}^{2}=-\left(  \gamma-\sigma\left(  nCr+B\right)  ^{\frac{1}{n}%
}\right)  dt^{2}+\frac{dr^{2}}{\left(  \gamma-\sigma\left(  nCr+B\right)
^{\frac{1}{n}}\right)  }+d\Sigma_{\gamma}^{2}~,
\end{equation}
where $\sigma=\left(  \pm1\right)  ^{n+1}$. Therefore in general the metric
distinguishes between the cases where the dimension is $d=4m$ or $d=4m+2$, and
a black hole is obtained when $\Sigma_{\gamma}$ is a manifold of negative
constant curvature. A detailed analysis of this kind of objects as well as
their thermodynamics is an interesting open problem.

We have considered so far the class of theories given by the action
(\ref{actionk}) which possesses the special property of having a single
maximally symmetric vacuum which is flat \cite{BH-Scan}. These theories were
obtained from the vanishing cosmological constant limit of the class of
theories with an action given by
\begin{equation}
I_{k}=\kappa\int\sum_{n=0}^{k}c_{n}^{k}\mathcal{L}^{n}~,
\end{equation}
with the following choice of coefficients
\begin{equation}
c_{n}^{k}=\left\{
\begin{array}
[c]{cc}%
\frac{l^{2\left(  n-k\right)  }}{\left(  D-2n\right)  }\left(
\begin{array}
[c]{c}%
k\\
n
\end{array}
\right)  & ,\text{ }n\leq k\\
0 & ,\text{ }n>k
\end{array}
\right.
\end{equation}
and $1\leq k\leq\left[  \frac{D-1}{2}\right]  $. This class of theories
possesses a unique maximally symmetric AdS vacuum with radius $l$. The de
Sitter case is obtained making $l\rightarrow il$. Exact spherically symmetric
black hole solutions for these class of theories exist, and their
generalization to the case with topologically nontrivial AdS asymptotics was
done in \cite{topbhscan}.

Extended black objects within these theories can be obtained from the
following class of metrics%
\begin{equation}
ds_{D}^{2}=f^{2}\left(  z\right)  d\tilde{s}_{D-1}^{2}+dz^{2}\text{~,}%
\end{equation}
where $d\tilde{s}_{D-1}^{2}$ can be any solution of the same
theory (as the ones found in \cite{BH-Scan}, \cite{topbhscan}) in
one dimension below
provided that the warp factor has the form%
\begin{equation}
f\left(  z\right)  =A\sinh\left(  \sqrt{\lambda}\left(  z-z_{0}\right)
\right)  ~,
\end{equation}
where $\lambda$ is determined by the $D$-dimensional cosmological
constant, and $A$ is related to the cosmological constant of the
same theory in $\left( D-1\right)  $ dimensions. When the
cosmological constant of the theory in $\left(  D-1\right)
$-dimensions tends to zero the warp factor tends to
$\exp(\sqrt{\lambda}\left(  z-z_{0}\right)  )$.

The thermodynamics of this class of extended objects is also an open problem.

\bigskip

\textbf{Acknowledgements:} We thank M. Aiello, A. Anabal\'{o}n, E.
Ay\'{o}n-Beato, G. Dotti, R. Gleiser, and S. Willison for helpful
comments. G. Giribet would like to express his gratitude to Centro
de Estudios Cient\'{\i}ficos (CECS), Valdivia, for the
hospitality, where this work was done. Besides, he thanks Natxo
Alonso-Alberca for several enjoyable discussions on related
topics. J. Oliva and R. Troncoso thank G. Giribet and R. Ferraro
for the hospitality at the Universidad de Buenos Aires (UBA), and
at the Instituto de Astronomia y Fisica del Espacio (IAFE) where
part of this work was done, and specially thanks G. Dotti and R.
Gleiser for enlightening discussions. J. Oliva is very grateful to
the project MECESUP UCO-0209. This work was partially funded by
CONICYT/SECYT CH/PA03-EIII/014 and FONDECYT grants grants 1040921,
1051056, 1061291, and by CONICET. The generous support to CECS by
Empresas CMPC is also acknowledged. CECS is a Millennium Science
Institute and is funded in part by grants from Fundaci\'{o}n Andes
and the Tinker Foundation.

\end{document}